\documentclass[epj-spec]{svjour}
\usepackage{graphics}
\DeclareSymbolFont{UPM}{U}{eur}{m}{n}
\SetSymbolFont{UPM}{bold}{U}{eur}{b}{n}
\DeclareMathSymbol{\upi}{0}{UPM}{25}
\def\tenG #1{\mbox{\boldmath $#1$}}
\def\tpi{{\tenG\upi}}
\def\ten  #1{\bf#1}
\def\tq{{\ten q}}
\newcommand{\eps}{\varepsilon}
\newcommand{\dd} {\rm d}

\begin{document}
\title{Generalized Entropy and Transport Coefficients of Hadronic Matter}
\author{Azwinndini Muronga\inst{1,2}\fnmsep\thanks{\email{Azwinndini.Muronga@uct.ac.za}}}
\institute{Institute for Theoretical Physics and Astrophyiscs,
  Department of Physics, University of Cape Town, Rondebosch 7701,
  South Africa \and UCT-CERN Research Centre, Department of Physics, 
  University of Cape Town, Rondebosch 7701, South Africa  }
\abstract{
We use the generalized entropy four-current of the
M\"uller-Israel-Stewart (MIS) theories of relativistic dissipative
fluids to obtain information about fluctuations around equilibrium
states. This allows one to compute the non-classical
coefficients of the entropy 4-flux in terms of the equilibrium
distribution functions. The Green-Kubo formulae are used to compute
the standard transport coefficients from the fluctuations of entropy
due to dissipative fluxes.
} 
\maketitle
\section{Introduction}
\label{intro}
Transport coefficients, such as viscosities, diffusivities, and
conductivities characterizes the dynamics of fluctuations in a
system. Transport phenomenon in relativistic fluids are of great
interest in connection with the study of astrophysical conditions such
as those in neutron stars, the study of the cosmological conditions
such as those in the early universe, as well as the study of the relativistic
nuclear collisions such as those at the Relativistic Heavy Ion
Collider (RHIC) and at the Large Hadron Collider (LHC).

Knowledge of transport coefficients and associated lengths and/or
time scales is important in comparing observables with theoretical
predictions. In the study of relativistic nuclear collisions knowledge of
transport coefficients will greatly advance our current efforts and
interests in the use of relativistic dissipative fluid dynamics in
describing the observables \cite{Muronga04I,MR,Heinz,Romatschke}.
 Relativistic heavy ion collisions offer the opportunity
to study interactions between hadrons over a wide range of net baryon
density (or baryon chemical potential) and energy density (or
temperature). In order to study the transport properties of hadronic
matter produced in such collisions one should extract the transport
coefficients and associated length/time scales for a given model of
interacting hadrons. Then using relativistic dissipative fluid dynamics 
one can study the sensitivity of the
space-time evolution of the system and the calculated distributions of
the hadrons to dissipative, non-equilibrium processes. In the end one
needs to compare the predicted distributions with those observed in
experiments.

\section{Second-order entropy 4-current in MIS theories}
\label{sec:entropy-4-current}

In the MIS theories of relativistic fluid dynamics the entropy 4-current is a
function of both the classical variables and dissipative fluxes, i.e.,
$S^\mu = s^\mu(\eps,n,u^\mu,\Pi,q^\mu,\pi^{\mu\nu})$ where $\eps$ is
the energy density, $n$ is the net charge, $u^\mu$ is the 4-velocity,
$\Pi$ is the bulk viscous pressure, $q^\mu$ is the heat flux and
$\pi^{\mu\nu}$ is the shear viscous pressure tensor.
The second order or generalized entropy 4-current in the MIS theories of
relativistic dissipative fluids \cite{MIS} 
may be written as \cite{Muronga07I,Muronga07II}
\begin{eqnarray}
S^\mu &=& s_{\rm eq} u^\mu +\beta q^\mu -{1\over 2}\beta u^\mu
\left(\beta_0 \Pi^2 - \beta_1 q^\nu q_\nu
  +\beta_2\pi^{\nu\lambda}\pi_{\nu\lambda}\right) \nonumber\\
&&-\beta\left(\alpha_0q^\mu \Pi - \alpha_1q_\nu\pi^{\mu\nu}\right) \label{eq:entro-4-current}
\end{eqnarray}
where $s_{\rm eq}(\varepsilon,n)$ is the equilibrium entropy
density, $u^\mu$ is the hydrodynamical 4-velocity of the net charge
  and is to be normalized such that $u^\mu u_\mu =1$, $\beta \equiv
  1/T$ is the inverse temperature. The non-classical coefficients
  $\alpha_i (\varepsilon,n)$ and $\beta_i (\varepsilon,n)$ in
  Eq. (\ref{eq:entro-4-current}) are expressed in terms of
  thermodynamic integrals or can be obtained as differentiations of
  the equation of state $p \equiv p(\alpha,\beta)$ with respect to
  $\alpha$ and $\beta$ (where $\alpha\equiv \beta\mu$ with $\mu$  the
  chemical potential). The transition to $p\equiv p(\varepsilon,n)$
  can be done through the relations
\begin{equation}
n =\beta{\partial p\over \partial \alpha}\,, \;\;\;\;\; \varepsilon =
-\left(p+\beta{\partial p \over \partial \beta}\right)~.
\end{equation}
These non-classical second order coefficients are presented in detail in
Ref. \cite{Muronga07II}.
\begin{figure}
\resizebox{0.75\columnwidth}{!}{%
  \includegraphics{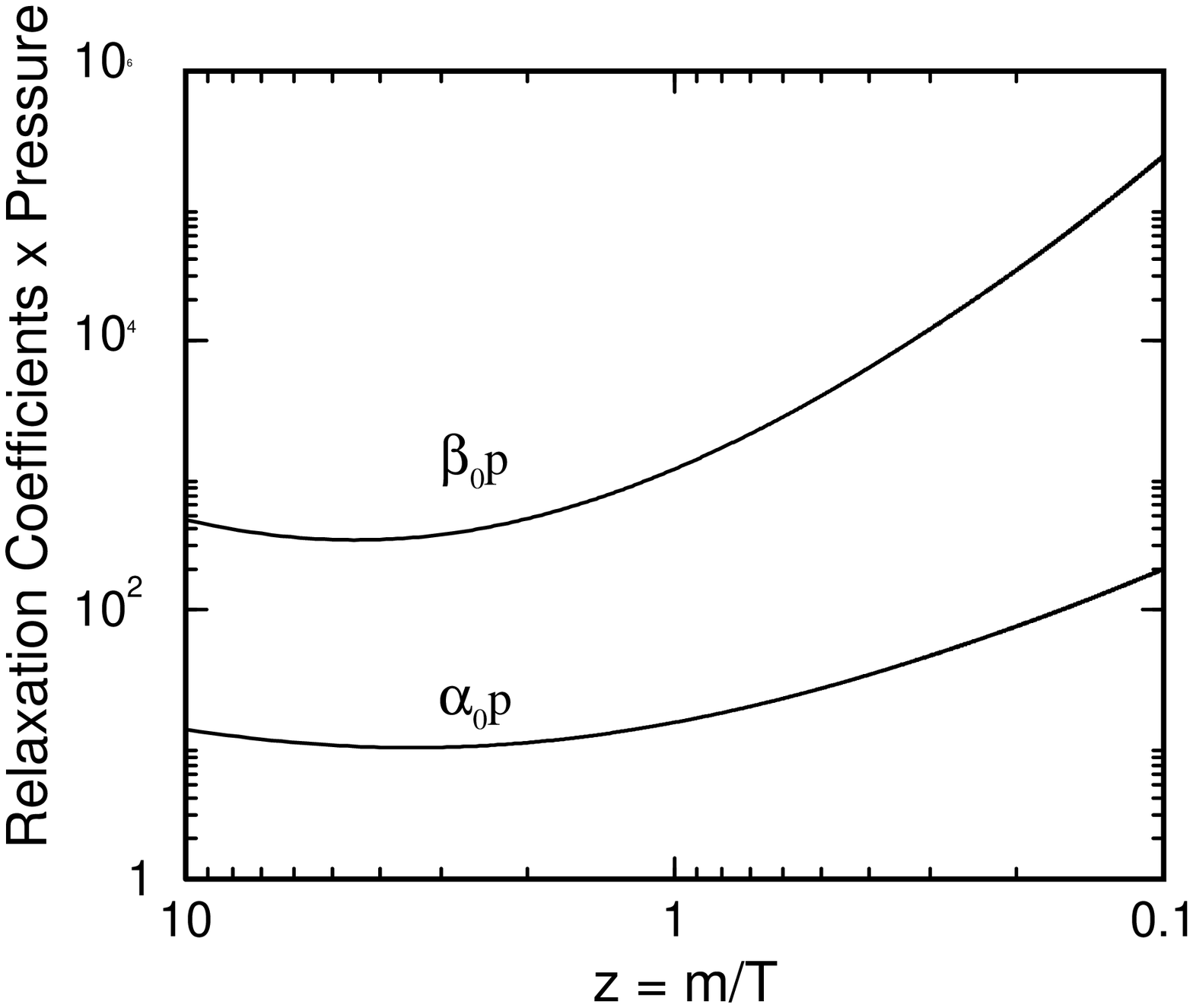} }
\caption{The temperature dependence of the  relaxation/coupling
  coefficients, $\alpha_0$ and $\beta_0$ for a pion gas.}
\label{fig:b0a0}       
\end{figure}
From the entropy 4-current, Eq. (\ref{eq:entro-4-current}), the
entropy density and flux are, respectively, given by
\begin{eqnarray}
s &=& u_\mu S^\mu = s_{\rm eq}(\varepsilon,n) -{1\over 2}
\beta\left(\beta_0\Pi^2 -\beta_1q_\nu q_\nu
  +\beta_2\pi^{\nu\lambda}\pi_{\nu\lambda}\right)~, \label{eq:entro-density}\\
\Phi^\mu &=& \Delta^{\mu\nu} S_\nu = \beta q^\mu - 
\beta(\left(\alpha_0q^\mu \Pi - \alpha_1q_\nu \pi^{\mu\nu}\right)~,\label{eq:entro-flux}
\end{eqnarray}
where $\Delta^{\mu\nu}=g^{\mu\nu}-u^\mu u^\nu$ is the projection tensor
with $g^{\mu\nu}$ = diag$(1,-1,-1,-1)$ the Minkowski metric tensor. 
The entropy density $s\equiv s(\varepsilon,n,u^\mu, q^\mu,
\Pi,\pi^{\mu\nu})$ is independent of the $\alpha_i$ while the entropy
flux is independent of $\beta_i$. The thermodynamic coefficients
$\beta_i(\varepsilon,n)\geq 0$ in Eq. (\ref{eq:entro-density}) model
deviations of the physical entropy density $s$ due to dissipative
contributions to $S^\mu$. The $\alpha_i(\varepsilon,n)$ in
Eq. (\ref{eq:entro-flux}) model contributions due to viscosity/heat
couplings, which do not influence the physical entropy.

\begin{figure}
\resizebox{0.75\columnwidth}{!}{%
  \includegraphics{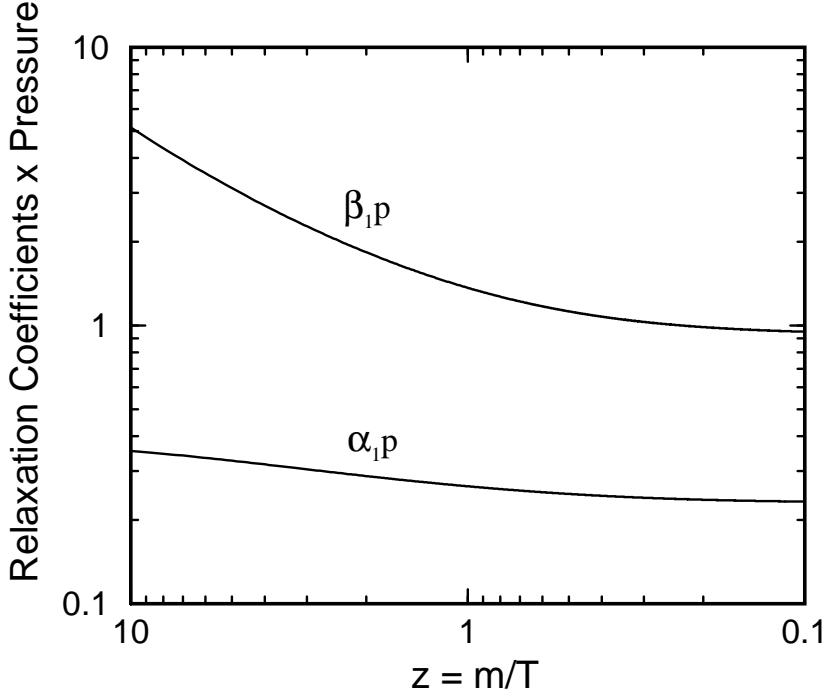} }
\caption{The temperature dependence of the  relaxation/coupling coefficients, 
 $\alpha_1$ and $\beta_1$, for a pion gas.}
\label{fig:b1a1}      
\end{figure}

\begin {figure}
\resizebox{0.75\columnwidth}{!}{%
  \includegraphics{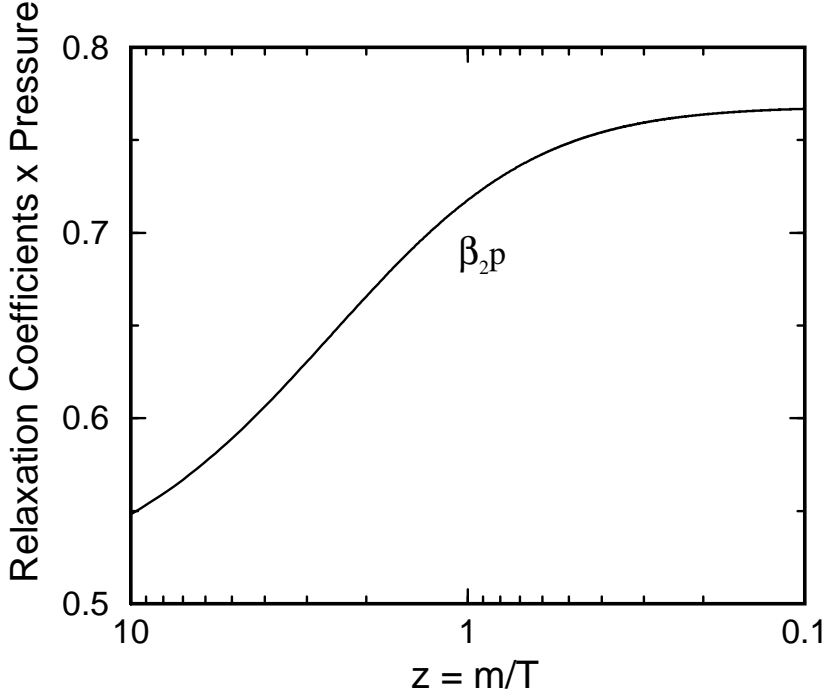} }
\caption{The temperature dependence of the shear viscous relaxation coefficient, 
 $\beta_2$, for a pion gas.}
\label{fig:b2}       
\end{figure}

The non-classical coefficients $\alpha_i$ and $\beta_i$ are shown in
Figs. \ref{fig:b0a0}, \ref{fig:b1a1} and \ref{fig:b2}. Shown in these
figures are these coefficients, multiplied by pressure so that they are
dimensionless, as functions of the ratio of pion mass to temperature
$z=m/T$. These non-classical coefficients are not
parameters but they are constrained by the equation of state. Thus to
compute the entropy one needs just the knowledge of either the
standard transport coefficients (in the case of a simple one component
fluid these are the bulk and shear viscosities and
the heat conductivity), or the corresponding relaxation times
to determine the dissipative fluxes.

\section{Second moments of equilibrium fluctuations}
\label{sec:2}

We shall now use the generalized entropy discussed in the previous
section to obtain the transport coefficients (cf. \cite{Jou}). 
In the local rest frame of the 3+1 formulation \cite{Muronga07I} 
the entropy density may be written as
\begin{equation}
s(\varepsilon,n,\tq,\Pi,\tpi) = s_{\rm eq}(\varepsilon,n)
- {1\over 2}\beta\left(\beta_0 \Pi^2 + \beta_1 \tq\cdot\tq
  +\beta_2\tpi:\tpi\right)
\end{equation}
The generalized Gibbs equation then takes the form
\begin{equation}
{\rm d} s = {\partial s\over \partial \eps} {\dd \eps} +{\partial
  s\over \partial n} \dd n + {\partial s\over \partial \tq}\cdot \dd
\tq + {\partial s\over \partial \Pi} \dd \Pi + {\partial s \over
  \partial \tpi} :\dd \tpi \label{eq:Gibbs}
\end{equation}
The probability $W$ of fluctuations of thermodynamic variables with
respect to their equilibrium reference values is related to the second
differential of the entropy by the Einstein formula \cite{Einstein} for 
probability of fluctuations
\begin{equation}
W \approx \exp\left[{1\over 2} \delta^2 S\right]~, \label{eq:Einstein}
\end{equation}
where $S$ is the total entropy of the system in volume $V$. For small
fluctuations around equilibrium states the change in entropy $\Delta S
= S - S_{\rm eq}$ can be expanded as 
\begin{equation}
\Delta S \approx (\delta S)_{\rm eq} + {1\over 2}(\delta^2 S)_{\rm eq}.
\end{equation}
Since the entropy is maximum in equilibrium, $(\delta S)_{\rm eq}$ =0
and $(\delta^2 S)_{\rm eq} \leq 0$. 
The second differential of the generalized entropy may be derived from the Gibbs
equation, Eq. (\ref{eq:Gibbs}), 
and when the resulting expression is introduced
into Eq. (\ref{eq:Einstein}) one obtains
\begin{eqnarray}
W(\delta \eps,\delta n,\delta \Pi,\delta \tq, \delta \tpi) &\approx& 
\exp \Bigl\{{1\over 2} V \Bigl[{\partial^2 s_{\rm eq}\over \partial \eps^2} (\delta
\eps)^2 +2 {\partial^2 s_{\rm eq}\over \partial \eps \partial n} \delta \eps \delta n +
{\partial^2 s_{\rm eq}\over \partial n^2} (\delta n)^2 
\Bigr.\Bigr. \nonumber\\
&&
\Bigl. \Bigl.~~~~~~~~~~~~~~~
-\beta\left(\beta_0\delta \Pi \delta \Pi + \beta_1\delta \tq\cdot \delta \tq + \beta_2
\delta \tpi :\delta \tpi \right) \Bigr]\Bigr\}
\end{eqnarray}
The second moments of a multi-variant distribution function 
\begin{equation}
W \approx \exp\left[-{1\over 2}M_{ij} \delta x_i \delta x_j\right]~,
\end{equation}
where $M_{ij}$ is the matrix corresponding to the second derivatives
of the entropy with respect to the classical and flux variables,
  are given by
\begin{equation}
\langle \delta x_i \delta x_j \rangle = (M^{-1})_{ij}, 
\end{equation}
where the brackets $\langle ... \rangle$ denote the average over the probability
distribution. In equilibrium the fluctuations of the classical variables $\eps$
and $n$ are uncoupled with the fluctuations of the fluxes. The second moments of
the fluctuations of $\Pi$, $\tq$ and $\tpi$  are given by
\begin{eqnarray}
\langle \delta \Pi \delta \Pi \rangle &=& {T\over \beta_0 V}~,\nonumber\\
\langle \delta q_i \delta q_j \rangle &=& {T\over \beta_1
V}\delta_{ij}~,\label{eq:fluctuations}\\
\langle \delta \pi_{ij} \delta \pi_{kl} \rangle &=& {T\over \beta_2
V}\Delta_{ijkl}~,\nonumber
\end{eqnarray}
with $\delta_{ij}$ the Kronecker symbol and
$\Delta_{ijkl}=(\delta_{ik}\delta_{jl}+\delta_{il}\delta_{jk} - {2\over 3}
\delta_{ij}\delta_{kl})$.

Recall that the classical coefficients of bulk viscosity $\zeta$, thermal
conductivity $\kappa$ and shear viscosity $\eta$ are related to the second order
non-classical coefficients $\beta_i$ by \cite{Muronga07II}
$\beta_0 = \tau_\Pi/ \zeta$, $\beta_1 = \tau_q/ \kappa T$ and $
\beta_2 = \tau_\pi/ 2\eta$ where the $\tau_A$ are the relaxation times for
the respective dissipative fluxes. Thus the expressions    
Eqs. (\ref{eq:fluctuations}) relate the dissipative coefficients $\zeta$,
$\kappa$ and $\eta$ to the fluctuations of the fluxes with respect to
equilibrium. Hence these coefficients determine the strength of the fluctuations
or vise versa the fluctuations determine the dissipative coefficients. The
expressions, Eqs. (\ref{eq:fluctuations}) for the second moments of the
fluctuations allow us to compute the dissipative coefficients of the
generalized entropy expression from the equilibrium distribution function and
microscopic expressions of the dissipative fluxes.

The Green-Kubo relations for bulk viscosity, thermal conductivity and shear
viscosity can be written as
\begin{eqnarray}
\zeta &=&  {V \over T}\int_0^\infty \langle \delta \Pi(0) \delta \Pi(t)\rangle dt ~,\nonumber\\
\kappa &=& {V \over T^2}\int_0^\infty \langle \delta q_i(0) \delta q_j(t)\rangle
dt ~,\label{eq:coeffs}\\
\eta &=& {V \over T}\int_0^\infty \langle \delta \pi_{ij}(0) \delta \pi_{ij}(t)\rangle
dt ~ \;\;\;\;\; (i \neq j) ~,\nonumber
\end{eqnarray}
respectively. In Eqs. (\ref{eq:coeffs}) the brackets $\langle ... \rangle$ stand
for equilibrium average, and no summation is implied over the repeated indices.
The time correlation functions can be calculated from
Eq. (\ref{eq:fluctuations}) together with Maxwell-Cattaneo relations
\cite {MC} as presented in Ref. \cite{Muronga07I} 
to give 
\begin{eqnarray}
\langle\delta\Pi(0) \delta\Pi(t)\rangle &=&\zeta T(\tau_\Pi V)^{-1}
\exp(-t/\tau_\Pi)\nonumber ~,\\ 
\langle\delta q_i(0) \delta q_j(t)\rangle &=&\lambda T^2(\tau_q V)^{-1} \delta_{ij} 
\exp(-t/\tau_q)
\label{eq:2ndmom} ~,\\
\langle\delta \pi_{ij}(0) \delta \pi_{kl}(t)\rangle &=&\eta T (\tau_\pi V)^{-1}
\Delta_{ijkl} \exp(-t/\tau_\pi) ~,\nonumber
\end{eqnarray}
If the evolution of the fluctuations of the fluxes is described by
Maxwell-Cattaneo relaxation equations (cf \cite{Muronga07II}), then after
integration, Eqs. (\ref{eq:coeffs}) reduce to
\begin{eqnarray}
\zeta &=& \tau_\Pi {V\over T} \langle \delta \Pi(0) \delta \Pi(0)\rangle\\
\kappa &=& \tau_q {V\over T^2} \langle \delta q_i(0)\delta q_i(0)\rangle ~,\\
\eta   &=& \tau_\pi {V\over T} \langle \delta \pi_{ij}(0) \delta
\pi_{ij}(0)\rangle ~.
\end{eqnarray}

\section{Hadronic Matter}

In spite of their importance, transport coefficients of hot, dense
hadronic gases are still difficult to calculate from first
principles.  Progress in the study of hadronic matter transport
coefficients is very slow, and only a calculation of transport
coefficients in the variational method \cite{Prakash,Davesne} and
relaxation time approximation \cite{Gavin} has been done. 
Though these models can describe some aspects of the properties of the
hadronic matter, whether they are realistic enough or not is
unclear. Thus, we need to investigate the transport properties of
a hadron gas by using a microscopic model that includes realistic
interactions among hadrons. This is of great interest in relativistic
nuclear collision \cite{Muronga04II,Sasaki}. In this work, we adopt a relativistic
microscopic model, UrQMD \cite{UrQMD} \ and perform
molecular--dynamical simulations for a hadronic gas of mesons in a box
of volume $V$.

We focus on the hadronic scale temperature ($100$ MeV $< T < 200$ MeV) and zero
baryon number density which are expected to be realized in the central high
energy nuclear collisions.  Transport coefficients of hadronic matter in this
region should play important roles in phenomenological models.  Sets of
statistical ensembles are prepared for the system at different 
energy densities. Using these ensembles, the shear viscosity coefficient
of a hadronic gas of mesons is studied as a function of temperature.

In computing the non-classical coefficients appearing in the
generalized entropy, Eq. (\ref{eq:entro-4-current}) we consider a
system of single component gas of pions. In Figs. \ref{fig:b0a0},
\ref{fig:b1a1} and \ref{fig:b2} we show the relaxation/coupling
coefficients for a hadronic gas of pions. The equation of state is
taken to be that of a resonance gas of non--interacting pions of mass,
$m_\pi=140$ MeV.  Knowledge of the above coefficients allows one to
write the primary coefficients in terms of the relaxation times. Such
relaxation times depend on the collision term in the Boltzmann
transport equation, and their derivation is an extremely laborious
task.

We now compute the the transport coefficient of shear viscosity of a meson gas
composed of $\pi,\eta,\omega,\rho$ and $\phi$. We use microscopic transport
model, the Ultra-relativistic Quantum Molecular dynamics (UrQMD) \cite{UrQMD}, 
using the Green-Kubo formulas. For detailed analysis see \cite{Muronga04II}.

\begin{figure}
\resizebox{0.75\columnwidth}{!}{%
  \includegraphics{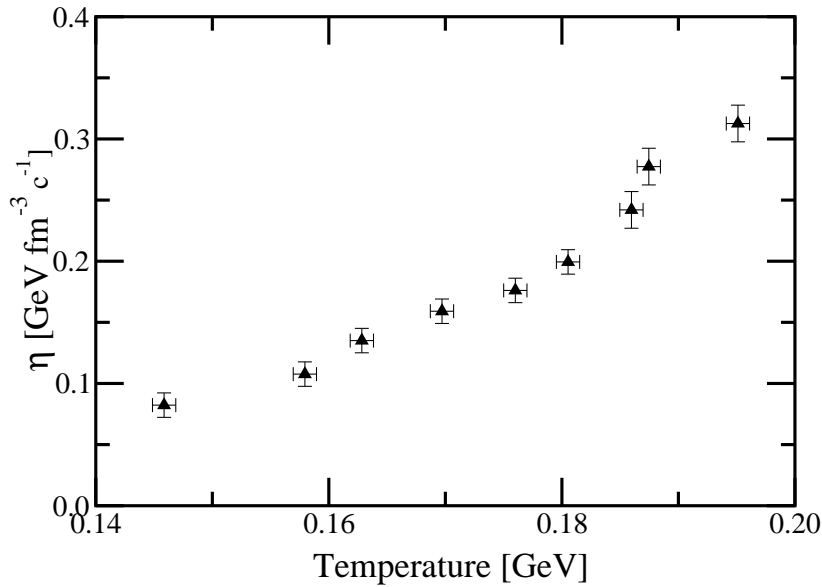} }
\caption{Shear viscosity of meson gas as a function of
temperature (cf. \cite{Muronga04II}).}
\label{fig:urqmdshear}       
\end{figure}

\begin{figure}
\resizebox{0.75\columnwidth}{!}{%
  \includegraphics{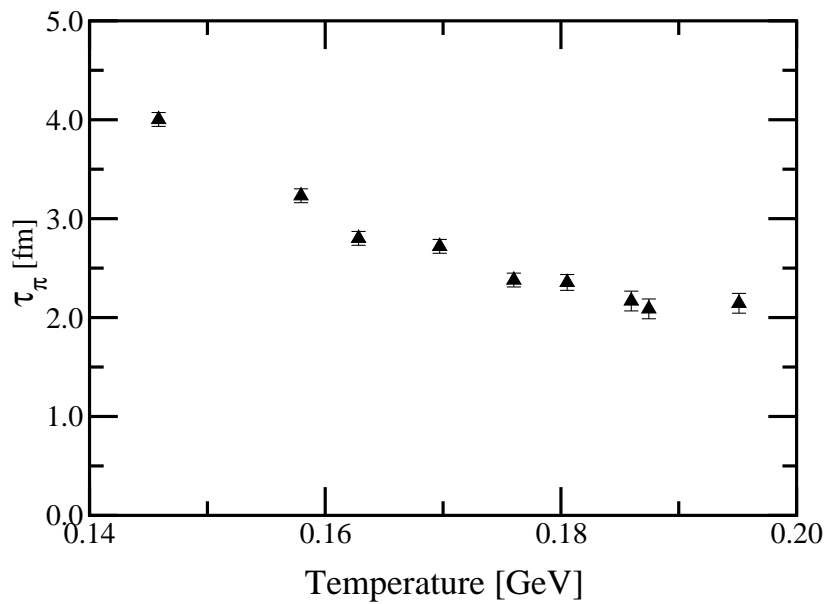} }
\caption{The relaxation time for the shear flux of meson 
gas as a function of temperature (cf. \cite{Muronga04II}).}
\label{fig:urqmdrelax}     
\end{figure}

Figure \ref{fig:urqmdshear} shows the shear viscosity coefficient results from
UrQMD using Kubo relations. As in the variational approach the coefficient grows
with temperature. The UrQMD results are about twice those from the variational
method. This might be due to the many meson resonances included in UrQMD while 
in the 
variational method we only have pions. Also the cross section parameterizations
are different in the two approaches. 
Figure \ref{fig:urqmdrelax} shows the relaxation time for shear flux in a hot pion
gas calculated from UrQMD by fitting the shear stress correlations. The
dependence of the shear relaxation time on temperature is similar to the one
obtained using variational method. The results obtained here are about a factor
of two less. The reasons are similar to the ones given above for the shear
viscosity coefficient.
The use of fluctuations through Kubo relations has the advantage of finding not
only the transport coefficients but also the corresponding relaxation times. In
addition it is also possible to obtain the relaxation coefficients such as
$\beta_2$. 

\section{Conclusion}

Using the generalized entropy 4-current and the fluctuation theory together
with the Green-Kubo formulas we presented the expressions for the standard
transport coefficients in terms of the correlations of the dissipative fluxes
and the associated relaxation times. The correlations of the dissipative fluxes
are related to the non-classical coefficients of the entropy 4-current. An
interesting point provided by this analysis is the reduction in the number of
independent parameters. In the simple one component relativistic causal fluid
dynamics \cite{Muronga07II} we have eight parameters, namely
$\tau_\Pi,\,\tau_q,\,\tau_\pi,\,\zeta,\,\kappa,\,\eta,\,l_{q\Pi},\,l_{q\pi}$.
Equations (\ref{eq:fluctuations}) together with two analogue expressions for
the correlations of the heat/viscous couplings provide five relations between
these parameters, so that we are left with just three independent parameters,
for instance $\tau_\Pi,\,\tau_q,\,\tau_\pi$. In this way the macroscopic
generalized entropy plus fluctuation theory give much more information than
the standard entropy 4-flux with the same number of free parameters.

The Green-Kubo relations are helpful in studying the evolution of the
dissipative fluxes. By combining these relations with the generalized entropy
expression provides an interesting point of contact between the macroscopic
and the microscopic approaches. This is also evident by our choice of using a
microscopic transport model, namely UrQMD in extracting the underlying
transport coefficients.

%

\end{document}